\documentclass[aps,twocolumn,floats,superscriptaddress,nopacs]{revtex4}
\usepackage{graphicx}
\usepackage{amssymb,amsmath}
\usepackage{color}
\usepackage{xcolor}
\usepackage{bm}
\usepackage{hyperref}
\begin{document}

\author{Aaram J. Kim}
\affiliation{Department of Physics, Kings College London, Strand, London WC2R 2LS, United Kingdom}

\author{Nikolay V. Prokof'ev}
\affiliation{Department of Physics, University of Massachusetts, Amherst, Massachusetts 01003, USA}

\author{Boris V. Svistunov}
\affiliation{Department of Physics, University of Massachusetts, Amherst, Massachusetts 01003, USA}
\affiliation{National Research Center Kurchatov Institute, 123182 Moscow, Russia}
\affiliation{Wilczek Quantum Center, School of Physics and Astronomy and T. D. Lee Institute, Shanghai Jiao Tong University, Shanghai 200240, China}

\author{Evgeny Kozik}
\affiliation{Department of Physics, Kings College London, Strand, London WC2R 2LS, United Kingdom}

\title{Homotopic Action: A Pathway to Convergent Diagrammatic Theories}

\date{\today}
\begin{abstract}
The major obstacle preventing Feynman diagrammatic expansions from accurately solving many-fermion systems in strongly correlated regimes is the series slow convergence or divergence problem.
Several techniques have been proposed to address this issue: series resummation by conformal mapping,
changing the nature of the starting point of the expansion by shifted action tools, and applying the homotopy analysis method to the Dyson-Schwinger equation. They emerge as dissimilar mathematical procedures aimed at different aspects of the problem. The proposed homotopic action offers a universal and systematic framework for unifying the existing---and generating new---methods and ideas to formulate a physical system in terms of a convergent diagrammatic series. It eliminates the need for resummation,  allows one to introduce effective interactions, enables a controlled ultraviolet regularization of continuous-space theories, and reduces the intrinsic polynomial complexity of the diagrammatic Monte Carlo method. We illustrate this approach by an application to the Hubbard model.
\end{abstract}

\maketitle

Methods of quantum field theory have underpinned remarkable breakthroughs in condensed matter physics for three quarters of a century~\cite{AGD}. They provide an arsenal of tools, based on Feynman diagrams, for systematic description of many-body correlations. Early on it was recognized that series of Feynman diagrams are not always meaningful if summed to high orders~\cite{Dyson1952}, but the versatility of constructing expansions around different starting points \cite{Hedin1965, Rubtsov2005, profumo2015, Rossi2016, Wu2017, Chen2019, Rossi2020}
and self-consistent renormalization of their building blocks for incorporating correlation effects in low orders \cite{Baym_Kadanoff, Hedin1965, AGD} have rendered the diagrammatic technique a widespread language of theoretical physics. Recent explosive development of algorithms for numeric summation of the series using stochastic sampling, the so-called diagrammatic Monte Carlo (DiagMC) approach \cite{Prokofev1998, Burovski2006,
Prokofev2007,VanHoucke2010,Kozik2010, VanHoucke2012, deng2015emergent, profumo2015, Iskakov2016df, Gukelberger2017df, Wu2017, Rossi2017cdet, moutenet2018determinant, Simkovic2019, Chen2019, Bertrand2019, Kim_PRL_2020, Vandelli2020db, Simkovic2020, Rossi2020, li2020diagrammatic}, has opened a new pathway to solving strongly correlated systems with high and \textit{a priori} controlled accuracy. The role of controlling and improving the series properties has become key for reaching this goal.

It was found, in particular, that convergence of the self-consistently renormalized (bold-line) technique~\cite{Baym_Kadanoff} with diagram order does not yet guarantee that the result is correct~\cite{Kozik2015}. On the other hand, a wide class of dressed diagrammatic expansions can be formulated as a Taylor series in the powers of a single parameter $\xi$, which has well-defined analytic properties free from misleading convergence~\cite{Rossi2016}. In this formalism, where the expansion is based on the so-called shifted action, the arbitrary choice of the zeroth-order action was shown to improve convergence~\cite{profumo2015, Wu2017, Chen2019,Rossi2020}. Furthermore, when a subset of strongly correlated degrees of freedom can be solved exactly, an expansion around this solution, accomplished, e.g., by diagrammatic extensions of the dynamical mean-field theory (DMFT)~\cite{Kusunose2006, Toschi2007, Rubtsov2008, Rubtsov2012, Rohringer2018rmp}, 
often has superior convergence properties~\cite{Gukelberger2017df, Vandelli2020db}. Whenever transforming the action does not yet yield a convergent series, a wealth of techniques, such as conformal mapping and numerous analytic continuation methods~\cite{Pollet2010, VanHoucke2012,profumo2015, Rossi2018,Simkovic2019, Bertrand2019}, allows us to reliably reconstruct the result behind the series by an \textit{a posteriori} protocol. Even certain cases with zero convergence radius become tractable~\cite{Pollet2010, Rossi2018}.

Nonetheless, formulating a many-body problem in terms of a convergent diagrammatic power series is important. It was demonstrated~\cite{Rossi2017ccp} that, when the series converge, the DiagMC approach circumvents the fundamental computational complexity of interacting fermions, known as the negative sign problem~\cite{loh1990sign, troyer2005sign}.
Since Feynman diagrams can be constructed directly in the thermodynamic limit, the only systematic error in the final answer comes from the truncation of the series at some large order $n$ and the truncation error drops exponentially with $n$ for a convergent series. Recent efficient algorithms based on summation of connected diagrams in terms of determinants \cite{Rossi2017cdet, moutenet2018determinant, Simkovic2019, Bertrand2019, Kim_PRL_2020, Rossi2020, li2020diagrammatic} take exponential in $n$ time to evaluate the order-$n$ sum, which implies only polynomial scaling of the calculation time with the inverse of the desired error bound. Moreover, fast convergence of the series is essential for novel DiagMC methods that compute real-time dynamic properties using symbolic integration~\cite{Taheridehkordi2019AMI, Vucicevic2020, Taheridehkordi2020grouping, Taheridehkordi2020susceptibility}, where fewer terms of the series could be obtained in principle. More generally, having to deal with the problem of reconstructing the answer from divergent series has been a major drawback of diagrammatic approaches, requiring additional expertise and labor, and impeding proliferation of DiagMC methods for nonexpert users.

In this Letter, we show that the shifted-action tools, conformal mappings, and homotopy ideas can be used to design what we propose to call a ``homotopic action," $S_h$. The diagrammatic expansions based on this action produce series that converge automatically in cases when conformal mapping in combination with the shifted action solves the problem, with guaranteed reduction of computational complexity and additional possibilities for further iterative refinements. We illustrate this idea by constructing a homotopic action for a prototypical fermionic system, the doped two-dimensional (2D) Hubbard model, in a challenging correlated regime where the standard diagrammatic expansion diverges. The guaranteed convergence of expansions based on $S_h$ allows direct evaluation of observables by DiagMC with a single parameter $n$ controlling the accuracy. To this end, we implement a DiagMC algorithm based on the connected determinant Monte Carlo (CDet)~\cite{Rossi2017cdet} method, and demonstrate that it substantially improves the accuracy of the result in comparison with that inferred from an analytic continuation of the original divergent series obtained by CDet method.

As an example of new capabilities naturally emerging in the homotopic action framework---and distinctively different from existing shifted-action and conformal-mapping approaches---we propose a protocol for ``anticollapse'' regularization of continuous-space theories. It solves, at least conceptually, the notorious problem of the zero convergence radius due to Dyson's collapse~\cite{Dyson1952} by generating a convergent expansion in terms of the bare coupling, such as, e.g., the Coulomb potential.

\begin{figure}[t!]
\centering
\includegraphics[width=0.75\columnwidth]{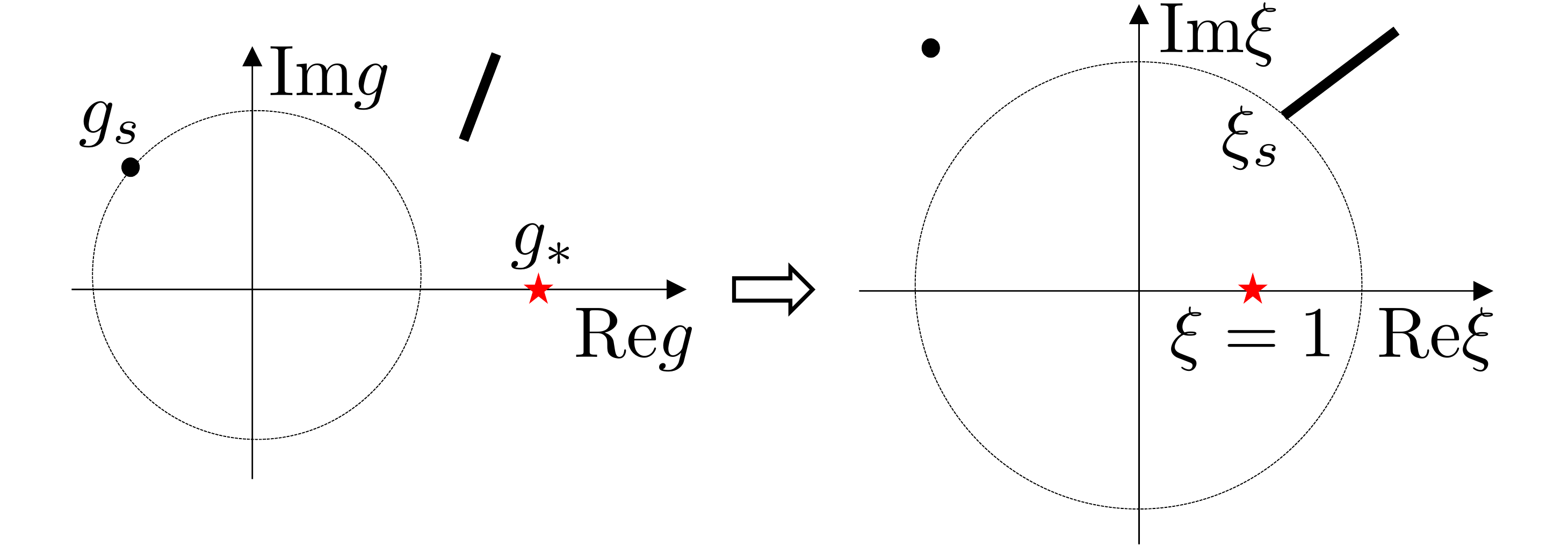}
\caption{Schematic of how a singularity $g_s$ in the complex plane of the coupling $g$ leads to a divergent series for the physical value $g_{*}$,
and how the shifted action trick works: it amounts to changing the
``origin of expansion" and introducing a different expansion parameter $\xi$;
the physical model is reproduced for $\xi=1$, the singularity $\xi_s$ controlling the convergence. Illustratory poles and branch cuts are depicted by dots and solid lines, respectively.
}
\label{fig1}
\end{figure}

{\it Shifted action as the simplest case of homotopic action}. A generic interacting fermionic system is described by an action of the form
\begin{equation}
S[\Psi] = S_0 [\Psi] + g S_{\rm int}[\Psi] \;,
\label{sa1}
\end{equation}
where $S_0$ is a bilinear in the Grassmann fields $\Psi$ part, $S_{\rm int}[\Psi]$ contains higher order in $\Psi$ interaction terms, and $g$ is the coupling constant. In strongly correlated regimes, one often finds that the most straightforward
approach to constructing Feynman diagrams---by expanding around $S_0$ in the powers of $g$---fails because the series diverge for the physical value of interest $g=g_{*}$. The convergence radius in the complex plane of $g$ can even be zero for models formulated in continuous space~\cite{Dyson1952}.
Unless the system undergoes a phase transition when $g$ is continuously increased from zero to $g_{*}$, this divergence stems from singularities in the complex plane, as illustrated in Fig.~\ref{fig1}, the closest one to the origin $g_s$ defining the convergence radius $|g_s|$.
Our intuition about such singularities is very limited because they are not necessarily
based on the ultraviolet physics or phase transitions taking place when the sign of $g$ is flipped, while for complex $g$ the Hamiltonian becomes non-Hermitian and hence unphysical.

 One way to get around the series divergence problem is to ``shift" the expansion point bringing the physics of interest inside the convergence radius (right half of Fig.~\ref{fig1}). In its simplest form, the idea~\cite{Hedin1965, profumo2015, Rossi2016, Wu2017} is to replace the original action (\ref{sa1}) with one of the form
\begin{equation}
\tilde{S}[\Psi; \xi] = \tilde{S}_0 [\Psi] + \Lambda [\Psi ; \xi]  + \xi  g S_{\rm int}[\Psi] \;.
\end{equation}
Here $\Lambda [\Psi ; \xi]$ is bilinear in $\Psi$ and it is assumed that its dependence on $\xi$
can be represented by a Taylor series
\begin{equation}
\Lambda = \sum_{j=1}^{\infty}\,  \xi^j \Lambda_j [\Psi] \;,
\label{sa2}
\end{equation}
convergent for any $\xi \le 1$. The only restriction that an arbitrary set of $\tilde{S}_0 $ and $\{ \Lambda_j \}$
functionals (the latter are called counter-terms) has to satisfy is
\begin{equation}
\tilde{S}_0 [\Psi] + \Lambda[\Psi ; \xi=1] =  S_0 [\Psi]  \;,
\label{sa3}
\end{equation}
so that $\tilde{S}(\xi=1)=S$.
The diagrammatic expansion is now performed in the powers of $\xi$ and $\tilde{S}_0$ serves as the new
state on top of which the expansion is done.

This tool can be used to expand around various mean-field and self-consistent solutions based
on a limited set of skeleton diagrams, such as the Hartree-Fock or \textit{GW} approximations, states with
explicitly broken symmetry, or any other approximate solution that is considered to be close
to the final answer. The resulting series in $\xi$ may happen to be convergent
even in the strongly correlated regime~\cite{profumo2015, Wu2017, Chen2019,Rossi2020}.

Going one step further, the two-body interaction terms between fermions
can be decoupled using the Hubbard-Stratonovich transformation involving complex-number fields $\varphi $
and the original action can be rewritten as
(in some cases this action is considered as the original one in the first place)
\begin{equation}
S[\Psi, \varphi] = S_0 [\Psi] + D_0 [\varphi] + \sqrt{g} V_{\rm int}[\Psi, \varphi] \;.
\label{sa4}
\end{equation}
At this point, the shifted action trick can be applied in both fermionic and bosonic channels
\begin{equation}
\tilde{S}(\xi) = \tilde{S}_0 [\Psi] + \tilde{D}_0 [\varphi] +
\Lambda [\Psi ; \xi] + \Omega [\varphi ; \xi] +  \sqrt{\xi g} V_{\rm int}[\Psi, \varphi] \;,
\label{sa5}
\end{equation}
and used to effectively change the nature of the interaction terms, order by order. The Taylor series in $\xi$ for $\Lambda$ and  $\Omega$ are chosen to
be convergent in the unit circle of $\xi$ [by construction, $\Lambda (\xi=0)=0$ and $\Omega (\xi=0)=0$]
with only one condition to satisfy for an otherwise infinite set of arbitrary functions:
$\tilde{S}(\xi=1) = S$.

Flexibility in designing shifted actions with a large number of non-linear in $\xi$ counterterms is
almost never used and most often only the simplest, linear in $\xi$, shifts $\Lambda$ of the Green's
function are implemented, with notable exception of Coulomb systems where screening
is required for having a meaningful expansion~\cite{Rossi2016,Chen2019}.
With the help of suitable Hubbard-Stratonovich transformations one can introduce new and manipulate
arbitrary many-body interactions~\cite{Rossi2016}, but the formalism is becoming progressively
more complex.

Beyond shifting, integrating out the original variables with the local action allows one to incorporate arbitrarily strong local correlations at the starting point of the expansion, as demonstrated by the dual fermion and boson theories~\cite{Rubtsov2008, Rubtsov2012, Rohringer2018rmp}. The dual diagrammatic series describing only nonlocal correlations are expected to be better behaved.

\begin{figure}[t!]
\centering
\includegraphics[width=1.0\columnwidth]{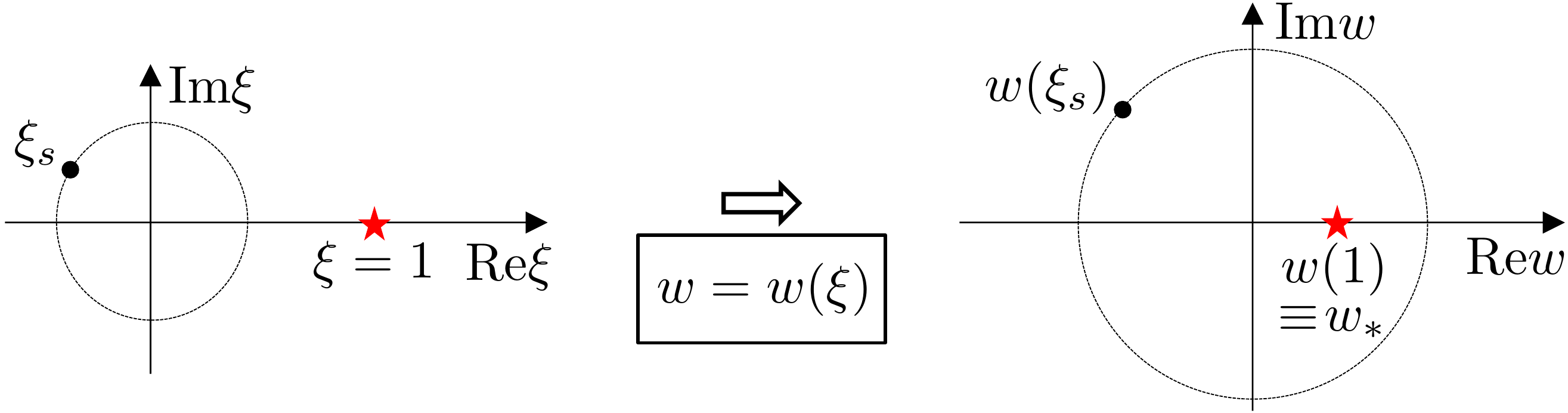}
\caption{Under conformal mapping, singularities defining the convergence radius (dot)
are moved farther away from the origin and the series for the point of interest (star) converge.
}
\label{fig2}
\end{figure}

Regardless of how the effective action is transformed, the resulting series may still diverge. So far, the standard protocol for dealing with this problem has been to apply series resummation techniques. The most versatile and widely used one is based on
conformal mapping, illustrated in Fig.~\ref{fig2}. First, one identifies singularities in
the complex plane of the expansion parameter, either analytically, using additional knowledge
about system properties \cite{Lipatov1977,Rossi2018, Bertrand2019}, or numerically, by matching the computed power series
$\sum_k a_k \xi^k$ by some analytic expression $A(\xi )$~\cite{Brezinski1996Pade, baker1961Dlog, hunter1979IA, mera2015hypergeom, mera2018maijer_g, Simkovic2019}: pole singularities are captured precisely by $A(\xi)$ that is a ratio of two polynomials, known as the Pad\'{e} approximant~\cite{Brezinski1996Pade}; brunch cuts are described by $A(\xi)$ of a more general form, e.g., the Dlog-Pad\'{e}~\cite{baker1961Dlog}, integral~\cite{hunter1979IA}, or hypergeometric and Meijer-$G$~\cite{mera2015hypergeom, mera2018maijer_g} functions. Knowing the location of the singularity $\xi_s$ closest to the origin, one can transform the complex plane of $\xi$ by an analytic function $w=w(\xi)$, $w(0)=0$, to a domain of the complex variable $w$ where the singularity is farther away from the origin than the image of $\xi=1$, $|w(\xi_s)|>|w(1)|$.
Then, upon expressing the inverse map $\xi(w)$ as
\begin{equation}
\xi = \sum_{k=1} f_k w^{k} \;,
\label{map}
\end{equation}
the reexpansion of the original series,
\begin{equation}
\sum_{j} a_j \xi^{j}(w) \longrightarrow  \sum_{k} b_k w^{k}(\xi) \;,
\label{map2}
\end{equation}
is convergent at $w(\xi=1)$. More generally, the mathematical literature on the topic of series resummation is vast, making divergent series practically useful, provided a sufficient number of terms is known and singularities are reasonably well understood. The homotopic action allows us to incorporate the principles of resummation in the formulation of the physical problem itself.

{\it General homotopic action}. The standard definition of homotopy is a continuous transformation of one function into another.
In the shifted action formalism described above, we aim at optimizing the diagrammatic expansion by selecting
an appropriate starting action $\tilde{S}(\xi=0)$ and its continuous transformation into the physical
action $S=\tilde{S}(\xi =1)$, similarly to the homotopy at the heart of the functional renormalization group (FRG)~\cite{salmhofer1999renormalization, berges2002fRG, fRG_RMP2012}, DMF$^2$RG~\cite{taranto2014DMF2RG}, and homotopy analysis~\cite{Pfeffer2017,Pfeffer2018} methods.
If we distance ourselves from the specifics of how various shifts are implemented,
we recognize that a far more
intuitive and transparent way to cast the attempted transformation of the action would
be to write
\begin{equation}
S_{h}(w) = \tilde{S}_0 + \Lambda_h (w) + \bar{S}_{\rm int} (w) \;,
\label{ha}
\end{equation}
where $\tilde{S}_0$ and $\Lambda_h$ are bilinear in all fields, the dependence of
$\Lambda_h (w)$ and $\bar{S}_{\rm int} (w)$ on $w$ can be represented by the convergent
Taylor series for $|w | \le |w_*|$, cf.~Eq.~(\ref{sa2}), and $S_{h}(w_*) \equiv S$ for some $w_*$.
There are no restrictions otherwise on the nature and number of terms contributing to
$\Lambda_h (w)$,  and  $\bar{S}_{\rm int} (w)$. They may be ``standard" counterterms based
on bare or skeleton diagrams, symmetry breaking and restoring fields, as well as 
{\it arbitrary} new interaction terms introduced by the homotopic transformation of the following 
(or similar) form:
\begin{equation}
S_{h}(w) \; \to \;  S_{h}(w) + w(w-w_*) S_{\rm eff} \, .
\label{h_eff}
\end{equation}
Here $S_{\rm eff}$ is chosen to capture the emerging physics of strong correlations already at the lowest orders of the expansion in $w$. 
Its form can be based on phenomenological considerations or explicit calculations in the framework of FRG~\cite{salmhofer1999renormalization, berges2002fRG, fRG_RMP2012} or DMF$^2$RG~\cite{taranto2014DMF2RG}.  

While it is hard to comprehend the ultimate potential of the homotopic action approach,
it is easy to see that it will automatically reproduce the result of conformal mapping, i.e. the diagrammatic series based on $S_{h}$ will converge. Indeed, if the series based on the action $\tilde{S}(\xi )$ diverge,
then Eq.~(\ref{map}) can be used to construct the homotopic action $S_{h}(w) \equiv \tilde{S}(\xi (w))$, with $w_*=w(\xi=1)$.
This procedure results in complete reshuffling of counterterms and interaction terms
between the orders in such a way that the expansion in the powers of $w$ is now convergent
because it is precisely the series on the rhs of (\ref{map2}), as follows from trivial power counting.
When this series is summed by DiagMC, the error bars on the final answer improve. The exploding in the limit of large $n$ original coefficients $a_n$ are combined into sign-cancelling and/or suppressed contributions to $b_n$. This leads to a reduction of the Monte Carlo variance, which is missing when the resummation (\ref{map2}) is applied after computing the coefficients $a_n$.This accuracy gain enables iterative improvement of the homotopic action $S_{h}(w)$ itself, aimed at further increasing the precision of the solution: Having analyzed the singularity structure in the $w$ plane, one could construct a subsequent conformal map $w'=w'(w)$ to obtain the action $S_{h}(w')=S_{h}(w(w'))$ with a faster-converging expansion, and so on.

On the practical side, there is no computational overhead in using $S_h$ instead of $\tilde{S}$ in DiagMC algorithms where the sum of all connected diagram topologies of a given order is performed deterministically using determinants and only the integration over the internal variables is done by Monte Carlo sampling~\cite{Rossi2017cdet, moutenet2018determinant, Simkovic2019, Kim_PRL_2020, Rossi2020, li2020diagrammatic} (the CDet algorithm and its derivatives). When computing the integrand of order $n$, these algorithms intrinsically evaluate contributions from all expansion orders up to $n$.
%
\begin{figure}[]
	\centering
	\includegraphics[width=0.8\columnwidth]{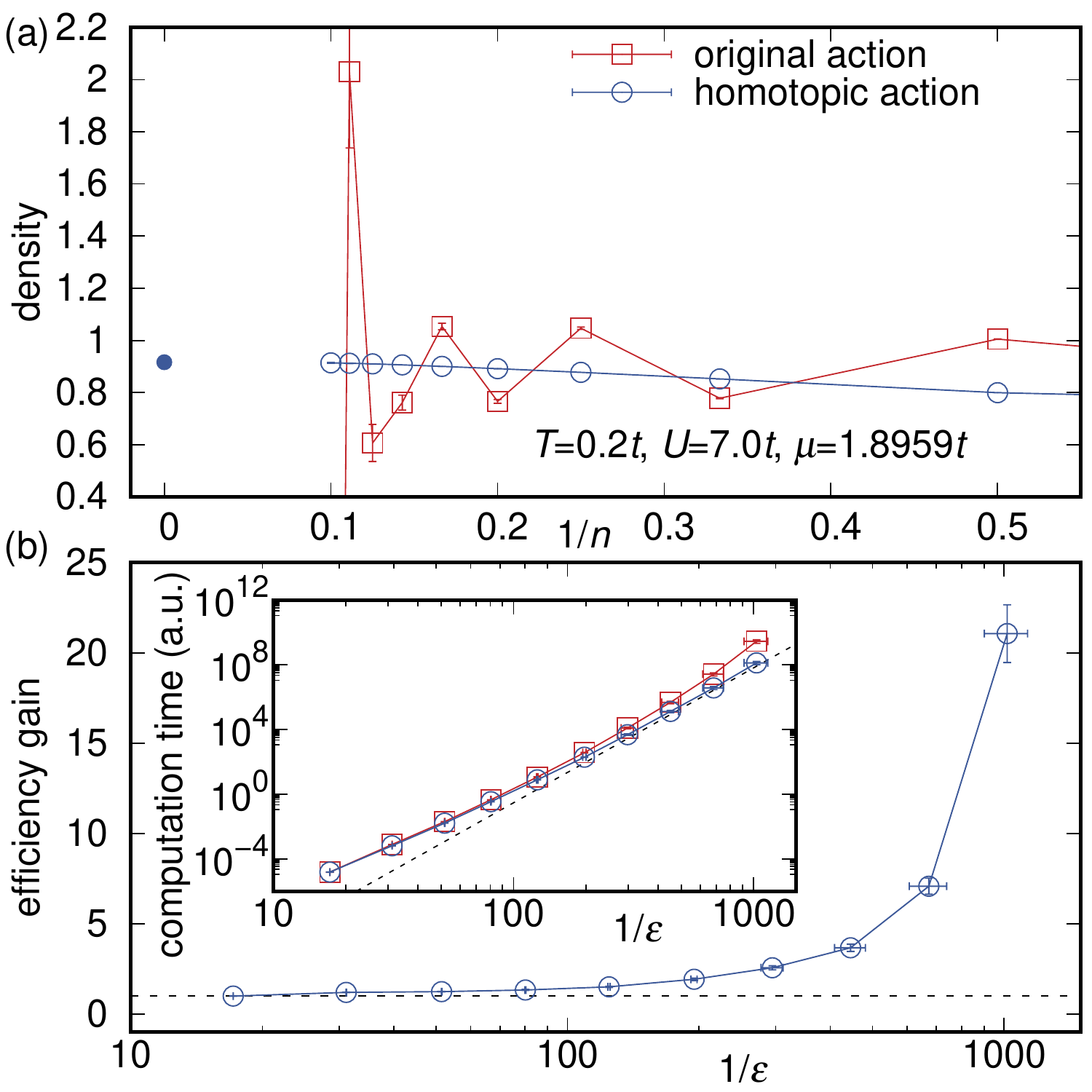}
	\caption{
		Results for the total fermion density of the 2D Hubbard model at $T=0.2t$, $U=7t$, $\mu=1.8959t$.
		(a) Partial sum of the divergent series for the original action $\tilde{S}(\xi=1)$ (\ref{eqn:Shm}) and the convergent series for the homotopic action $S_h(w_*)$.
		(b) Efficiency gain of $S_h(w)$ over $\tilde{S}(\xi)$ with its series resummed by $w(\xi)$, defined as the ratio of the respective computational times needed to obtain density within the error $\varepsilon$. Inset: the corresponding computational times and the reduction of the asymptotic polynomial scaling (dotted line).}
	\label{fig:density_error_ratio}
\end{figure}

Figure~\ref{fig:density_error_ratio} illustrates the efficiency of the homotopic action approach within the CDet framework by its application to the 2D Hubbard model on the square lattice. The standard shifted action for the model~\cite{Rubtsov2005, profumo2015, Wu2017} reads
\begin{eqnarray}
\tilde{S}\left[\Psi,\xi\right] = \tilde{S}_0\left[\Psi\right] + \Lambda [\Psi ; \xi] + \xi US_{\rm int}\left[\Psi\right]~,\nonumber\\
\tilde{S}_0\left[\Psi\right] = \sum^{}_{i\sigma}\int_{0}^{1/T}d\tau~\overline{\Psi}^{}_{i\sigma}(\tau)(\partial_\tau - \mu + \alpha)\Psi^{}_{i\sigma}(\tau)\nonumber\\
	- t\sum^{}_{\langle ij\rangle\sigma}\int_{0}^{1/T}d\tau~\left( \overline{\Psi}^{}_{i\sigma}(\tau) \Psi^{}_{j\sigma}(\tau) + c.c. \right)~,\nonumber\\
\Lambda [\Psi ; \xi] = - \xi \, \alpha~\sum^{}_{i\sigma}\int_{0}^{1/T}d\tau~~\overline{\Psi}^{}_{i\sigma}(\tau)\Psi^{}_{i\sigma}(\tau)~\;,\nonumber\\ 	
	S_{\rm int}\left[\Psi\right] =  \sum^{}_{i}\int_{0}^{1/T}d\tau \overline{\Psi}^{}_{i\uparrow}(\tau)\Psi^{}_{i\uparrow}(\tau)\overline{\Psi}^{}_{i\downarrow}(\tau)\Psi^{}_{i\downarrow}(\tau)~\;.\nonumber\\
	\label{eqn:Shm}
\end{eqnarray}
Here, $\Psi_{i\sigma}(\tau)$ represents the spin-$\sigma$ Grassmann field at the imaginary-time $\tau$ on the site $i$;
$T$ is the temperature, $\mu$ chemical potential, $t$ nearest-neighbor hopping amplitude, $U$ on-site repulsion, and $\alpha$ is the arbitrary shift parameter. Despite the shift, diagrammatic expansions with this action are divergent, e.g., for $T=0.2t$, $U=7t$, $\mu=1.8959t$, $\alpha=2.5568$ due to a singularity at $\xi_s \sim -0.65$. We construct the homotopic action by $S_{h}(w) \equiv \tilde{S}[\xi (w)]$, which generates a convergent expansion with $\xi(w) = 12w/7(1-w)^2$ mapping the branch cut along the real axis from $\xi=-3/7$ to $-\infty$ onto the unit circle $|w|=1$.

Figure~\ref{fig:density_error_ratio}(a) presents the partial sum of the convergent series $\sum_n b_n w_*^n$ for the total density with $b_n$ generated by the homotopic action $S_h$ contrasted to that of the divergent series $\sum_n a_n$ produced by the original action $\tilde{S}$.
The solid circle is the result of extrapolation of the convergent series to infinite order using the Dlog-Pad\'e method~\cite{baker1961Dlog}.
Note that, within the homotopic action framework, the Monte Carlo algorithm can directly sample the partial sum $\sum_{n}b_n w^n$ instead of the coefficient $b_n$ (see Supplemental Material~\cite{Kim:supp} for technical details).

When it comes to high-precision calculations, the homotopic action approach yields a significant
efficiency gain as compared to the conventional conformal mapping method based on postprocessing
of the original series [Fig.~\ref{fig:density_error_ratio}(b)]. The origin of this gain, as well as its sharp growth as a function of the inverse relative error, $1/\varepsilon$, within which the final result is obtained, is easy to understand. The gain is all about the way the high-order diagrams---important for achieving small $\varepsilon$---are sampled.
Instead of sampling diagrammatic contributions with large weights that nearly compensate in the final answer and leave one
with large relative error bars, the cancellation of the
sign-alternating terms from different
orders is now enforced before sampling; see~\cite{Kim:supp} for more details.


{\it Anticollapse regularization.} Dyson's reasoning~\cite{Dyson1952} for the zero convergence radius of perturbative expansions in continuous-space systems is directly linked to the ultraviolet (UV)
behavior of attractive fermionic fields: For an action (\ref{sa1}), an observable cannot be analytic at $g=0$ if changing the sign of $g$ for $|g| \to 0$ leads to a {\it collapse}---an instability towards unlimited increase of particle density.
In systems with hard momentum cutoff, e.g., lattice models, the expansion in the powers of coupling 
is expected (as supported by strong evidence) to have a finite convergence radius at $T>0$ . 
This observation leads to a natural homotopic procedure for constructing a theory with a controlled 
anticollapse UV regularization. The simplest trick (cf. infrared regularization in the FRG~\cite{salmhofer1999renormalization, berges2002fRG, fRG_RMP2012, taranto2014DMF2RG}) is to modify the free-particle dispersion,
\begin{equation}
\varepsilon (k)\; \to \; \varepsilon (k) +  \alpha (w_* - w) k^4  \;\;\;\; (\alpha>0) \;.
\label{k_to_4}
\end{equation}
At small $|w|$, the quartic term (\ref{k_to_4}) prevents Dyson's collapse, allowing one to explore analyticity in $w$
and extrapolation to the $w=w_*$ limit. A more advanced and general tool---a regularization of the interaction---is discussed in the Supplemental Material.


In conclusion, the paradigm of homotopic action $S_h(w)$, such that $S_h(w=0)$ is harmonic and $S_h(w=w_*)$ is identical to the physical action, reveals a broad family of convergent quantum-field-theoretical expansions
in the powers of a single (homotopy) parameter $w$. With an appropriately designed $S_h(w)$, one can naturally unify the shifted-action and resummation techniques, as was illustrated by a simple and yet nontrivial example. Further intriguing possibilities to explore in the future include, e.g. (i) ultraviolet (anticollapse) regularization of continuous-space theories, and (ii) the introduction in $S_h(w)$ of effective interactions that vanish both at $w=0$ and $w=w_*$ but are otherwise arbitrary and chosen to capture the physics of the model already at the lowest orders of expansion, as, e.g., in approximate analytic theories. As the expansion progresses, the homotopic action accomplishes a seamless replacement of the effective-interaction contributions by those from the original bare interaction, thereby establishing control of accuracy in effective theories.

\begin{acknowledgments}
E.K. is grateful to the Precision Many-Body Group at UMass Amherst, where a part of this work was carried out, for hospitality. This work was supported by EPSRC through Grant No. EP/P003052/1 (A.K. and E.K.) and by the Simons Collaboration on the Many-Electron Problem (N.P., B.S., and E.K.). N.P. and B.S. were supported by the National Science Foundation under the Grant No. DMR-1720465 and the MURI Program ``Advanced quantum materials -- a new frontier for ultracold atoms" from AFOSR.
We are grateful to the United Kingdom Materials and Molecular Modelling Hub for computational resources, which is partially funded by EPSRC (Grants No. EP/P020194/1 and EP/T022213/1).
\end{acknowledgments}

\bibliography{ref.bib}

\end{document}